\documentclass[doublecol]{epl2}

\usepackage{amsmath}
\usepackage{amsfonts}
\usepackage{graphicx}
\usepackage{dcolumn}
\usepackage{bm}
\usepackage{hyperref}
\usepackage{xcolor}
\usepackage[normalem]{ulem}

\title{Finite-Time Optimal Control by Noisy Traps}
\shorttitle{Finite-Time Optimal Control by Noisy Traps}

\author{Luca Cocconi\inst{1,2} \and Henry Alston\inst{3} \and Thibault Bertrand\inst{4}}
\shortauthor{L. Cocconi \etal}

\institute{
  \inst{1} Cavendish Laboratory, University of Cambridge, Cambridge CB3 0US, United Kingdom\\
  \inst{2} Max Planck Institute for Dynamics and Self-Organization (MPI-DS), 37077 G{\"o}ttingen, Germany\\
  \inst{3} {Laboratoire de physique de l'\'Ecole Normale Sup\'erieure,
		CNRS, PSL University, Sorbonne Universit\'e, and Universit\'e 
		Paris-Cit\'e, 75005 Paris, France}\\
  \inst{4} Department of Mathematics, Imperial College London, London SW7 2BZ, United Kingdom
}

\abstract{
The optimal control of passive systems in equilibrium typically favours quasistatic (infinite-time) protocols. We show that a breakdown of quasistatic optimality occurs when the controller itself is dissipative. Concretely, we study a Brownian particle confined by a harmonic trap with stochastically fluctuating stiffness, driven by an external protocol. When these fluctuations violate detailed balance, the probe-controller coupling continuously exchanges work with the system, altering the optimisation landscape. In this regime, optimal protocols are characterised by a finite duration which vanishes above a critical fluctuation strength. This transition can be directly observed in a short-time expansion of the mean work functional. When imposing an endpoint constraint, the transition to zero duration disappears and finite duration protocols remain optimal for all values of the controller fluctuations. These results demonstrate that finite-time optimality can emerge in passive systems under nonequilibrium control.
}

\newcommand{\M}{\mathsf{M}}
\newcommand{\du}{{\rm d}u}
\newcommand{\dw}{{\rm d}w}
\newcommand{\dt}{{\rm d}t}

\DeclareMathOperator*{\argmin}{arg\,min}

\begin{document}

\maketitle

\section{Introduction}

A central question in finite-time stochastic control is whether irreversibility can be reduced indefinitely by driving more slowly, or whether the controlled dynamics themselves select a non-trivial optimal duration. In the classic idealised setting of a passive Brownian probe manipulated by a deterministic harmonic potential, the answer is the former: while finite-time control is generically associated with nontrivial discontinuities in the optimal protocols \cite{schmiedl2007optimal}, the mean work simply approaches the free-energy difference in the quasistatic limit, and the excess dissipation can be made arbitrarily small by increasing the protocol duration \cite{Jarzynski1997}. 

By contrast, when irreversibility is sustained during the control itself --- for instance by internal activity of the probe, as in the case of self-propelled particles \cite{bechinger2016active,garcia2025optimal} or active field theories \cite{soriani2025control} --- quasistatic driving need no longer be globally optimal. In recent years, building on linear response theory \cite{sivak2012thermodynamic,zulkowski2012, Alvarado2026}, a geometric picture has emerged in the weak and slow driving regime which links finite time optimality in the control of active probes to a tradeoff between passive and active contributions, scaling asymptotically with the protocol duration $T$ respectively like $1/T$ and $T$ \cite{davis2024active,wang2025geometry}. These questions are also seeing renewed experimental interest \cite{blickle2012realization,tal2020experimental,saha2023information,khadka2018active,jun2014high,loos2024universal}, thanks to advances in the robust manipulation, particularly of colloids, at the micro-scale.

\begin{figure}
    \onefigure[width=\linewidth]{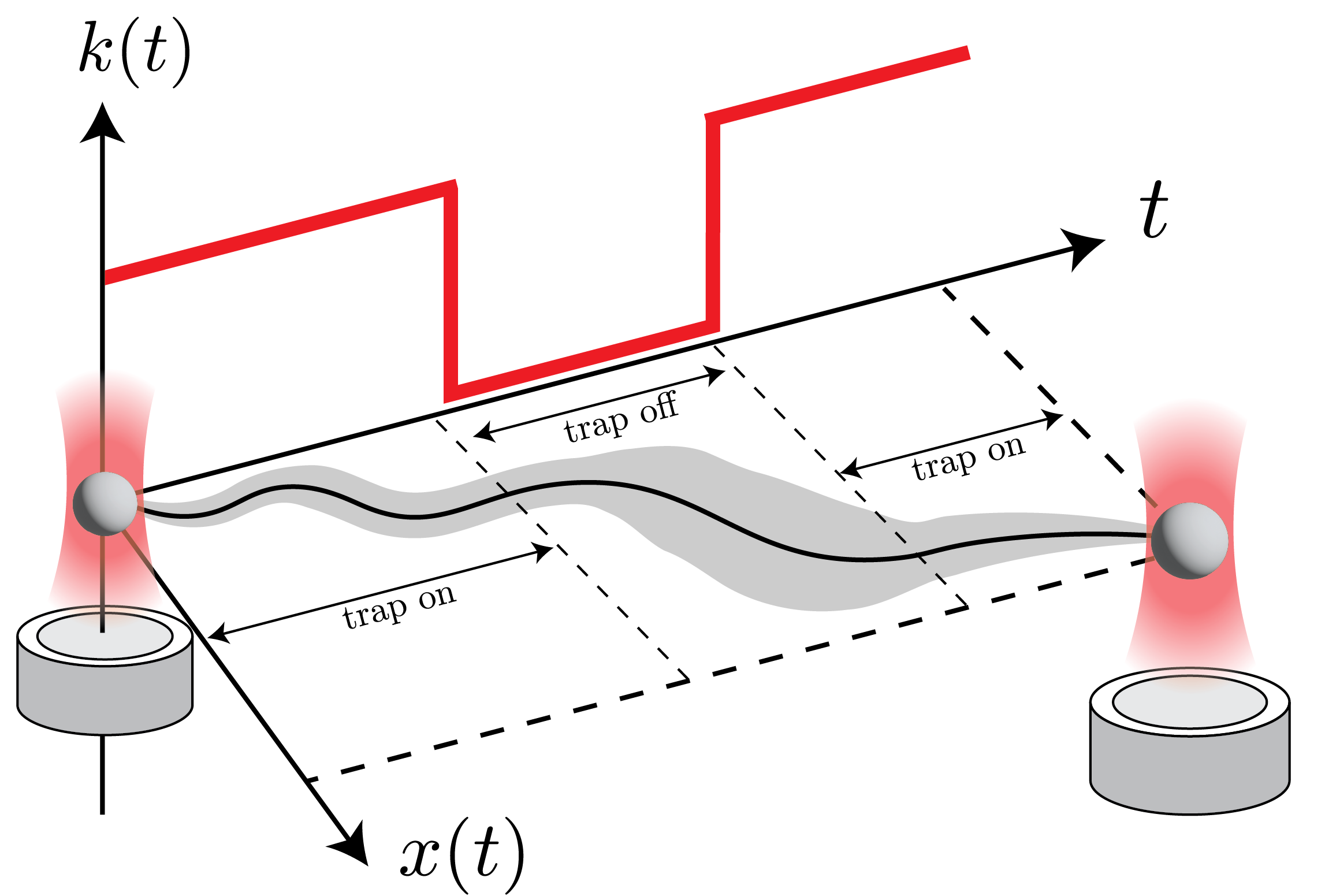}
    \caption{Schematic representation of a passive probe controlled by a noisy harmonic trap, here visualised as an intermittent optical tweezer. In the transport scenario, a deterministic protocol translates the trap centre along the coordinate $x$, while the trap stiffness $k(t)$ fluctuates stochastically and continuously exchanges work with the probe. 
    }
    \label{fig:schematic}
\end{figure}

Here we show that the breakdown of quasistatic optimality does not require an active probe. It already occurs for a passive Brownian particle provided that the \emph{controller} is dissipative. Concretely, we consider a harmonic trap whose stiffness $k(t)\in\mathcal{K}$ fluctuates stochastically while an externally prescribed control parameter $\lambda(t)$ drives either the trap centre or its mean stiffness. This setup, illustrated schematically in Fig.~\ref{fig:schematic}, is motivated both by recent theory on dissipative confinement \cite{cocconi2024ou2,alston2022non} and by experimental situations, particularly in the context of optical tweezers, in which the effective trapping potential is itself noisy or engineered to follow a prescribed noise model \cite{tal2020experimental,bustamante2021optical,pesce2020optical}. When the stiffness dynamics violate detailed balance \cite{peliti2021stochastic,liepelt2007kinesin} with respect to nonseparable energy terms coupling controller and probe, the joint process sustains stationary probability currents. This violation of time-reversal symmetry is signalled by a finite rate of entropy production \cite{cocconi2020entropy} even at fixed values of the deterministic control parameter $\lambda$ \cite{cocconi2024ou2,alston2022non}. 

We revisit, in this nonequilibrium setting, the two paradigmatic control problems analysed in Ref.~\cite{schmiedl2007optimal}: transport by moving the trap centre, and stiffening/softening by changing the trap stiffness. In both cases we find qualitative departures from equilibrium control. First, optimal protocols occur at finite duration even though the probe itself is passive. Second, above a critical level of stiffness fluctuations, the optimal duration of the unconstrained problem collapses to zero. This transition can be understood directly from the small-$T$ expansion of the optimised work: a positive term linear in $T$, generated by the noisy controller, competes with the short-time contribution from deterministic control, and the sign change of the linear coefficient marks the onset of local optimality of the zero-time protocol. In both scenarios, imposing a final-state constraint, which restricts the admissible control manifold, removes this transition. Finally, we show that the transport results are robust to temporally correlated forcing, although the zero-time transition is sensitive to the short-time structure of the noise source and disappears in this case. 

\section{Probe transport}\label{sec:transport}

First, we consider the case of a probe being transported by a translating harmonic trap with fluctuating stiffness. Namely, the center of the trap is moved dynamically and follows the path $\lambda(t)$. The probe position then obeys the overdamped Langevin dynamics
\begin{equation}\label{eq:langevin_transport}
    \dot{x}(t)=-k(t)\,[x(t)-\lambda(t)]+\sqrt{2D}\,\eta(t),
\end{equation}
where $\eta(t)$ is a zero-mean and unit-variance Gaussian white noise and $D$ is the diffusion coefficient. The stochastic stiffness $k(t)\in\mathcal{K}$ is modelled as a continuous-time Markov chain with transition-rate matrix $\M$. We denote by $\pi$ the stationary distribution of the chain, and by
\begin{equation}
    \bar{k}\equiv \sum_i k_i \pi_i,\qquad \sigma_k^2\equiv \sum_i (k_i-\bar{k})^2\pi_i
\end{equation}
its first two moments. 

To analyse the mean work, it is convenient to introduce the translational noise averages
\begin{align}
    u(t) \equiv \mathbb{E}_\eta[x(t)], \quad 
    w(t) \equiv \mathbb{E}_\eta[x^2(t)].
\end{align}
For a fixed realisation of $k(t)$, Eq.~\eqref{eq:langevin_transport} implies
\begin{subequations}\label{eq:langevin_u_w}
\begin{align}
    \dot{u}(t) &= -k(t)\,[u(t)-\lambda(t)],\\
    \dot{w}(t) &= -2k(t)\,[w(t)-\lambda(t)u(t)] + 2D.
\end{align}
\end{subequations}
Let $P_i(u,w,t)$ denote the joint probability density of $(u,w)$ and the stiffness mode $k(t)=k_i$. It obeys the Fokker-Planck equation
\begin{align}
    \partial_t P_i(u,w,t)
    &= -\partial_u\!\left[-k_i(u-\lambda)P_i\right]\nonumber\\
    &\quad -\partial_w\!\left[2k_i(\lambda u-w)P_i+2DP_i\right]+\M_{ij}P_j.
    \label{eq:FP_ot}
\end{align}
Summation over repeated indices is assumed throughout.

The stochastic work performed by the control upon application of a protocol $\lambda(t)$ of duration $T$ is 
\begin{equation}\label{eq:work_func_transport}
    W[\lambda] = \int_0^T \dt \ \dot{\lambda} k (\lambda-x) + \frac{\dot{k}}{2}(\lambda-x)^2~.
\end{equation}
Eq.~\eqref{eq:work_func_transport} differs from similar expressions, e.g., in Refs.~\cite{schmiedl2007optimal} and \cite{loos2024universal}, by the second term in the integrand, which vanishes for a deterministic potential.
After averaging over both thermal noise and stiffness fluctuations we define $\overline{W}[\lambda]\equiv \mathbb{E}_{\eta,k}[W[\lambda]]$, which is given by
\begin{equation}\label{eq:OT_def}
    \overline{W}[\lambda] \ =
     \int_0^T \dt \ \dot{\lambda}(\lambda \bar{k} - \langle k u \rangle) 
    + \frac{\lambda^2 \langle\dot{k} \rangle + \langle\dot{k} w\rangle - 2\lambda\langle\dot{k} u \rangle}{2} 
\end{equation}
Here angular brackets denote the average over the full joint process. In particular,
\begin{subequations}
\begin{align}
    \langle k f(u,w)\rangle &\equiv \sum_i \int \du \ \dw \ k_i f(u,w) P_i(u,w) \\
    \langle \dot{k} f(u,w)\rangle &\equiv \sum_{i,j} \int \du\ \dw\ f(u,w)(k_i - k_j)M_{ij} P_j(u,w)~. \label{eq:exp_invol_k}
\end{align}
\end{subequations}
For later use we also introduce the notation $\langle f(u,w)\rangle_i \equiv \int \du \int \dw \ f(u,w)P_i(u,w,t)$, such that full expectations may be written as $\langle f(u,w)\rangle = \sum_i \langle f(u,w)\rangle_i$.

\subsection{Dragging at constant speed}

Before turning to finite-time optimal transport, it is useful to inspect steady dragging with $\lambda(t)=vt$. 
In this case, one obtains the stationary work rate
\begin{equation}
    \mathbb{E}_{\eta,k}[\dot{W}_{\rm steady}]
    =v^2-D\bar{k}+\sum_i k_i\left[k_i \mu^{(2)}_{i}+v\,\mu^{(1)}_{i}\right],
    \label{eq:steady_drag_work}
\end{equation}
where the moments are defined by
\begin{subequations}
\begin{align}
    \mu^{(1)}_{i} &\equiv \lim_{t\to\infty}\langle u-vt\rangle_i, \\
    \mu^{(2)}_{i} &\equiv \lim_{t\to\infty}\langle w-2uvt+v^2t^2\rangle_i.
\end{align}
\end{subequations}
This follows by defining the displacement relative to the trap centre, $\tilde{x}=x-vt$ and combining the equivalent form $\mathbb{E}_{\eta,k}[\dot{W}_{\rm steady}]=-v\langle k\tilde{x}\rangle+\langle \dot{k}\tilde{x}^2\rangle/2$ with the stationary equations obtained by multiplying the Fokker-Planck equation for $P_i(\tilde{x})$ by $\tilde{x}$ and $\tilde{x}^2$ and subsequently integrating with respect to $\tilde{x}$. The resulting moments solve
\begin{subequations}
\begin{align}
    (\M_{ij}-k_i\delta_{ij})\mu^{(1)}_{j} &= v\,\pi_i, \label{eq:mom1}\\
    \left(\frac{\M_{ij}}{2}-k_i\delta_{ij}\right)\mu^{(2)}_{j} &= v\,\mu^{(1)}_{i}-D\pi_i.
    \label{eq:mom2}
\end{align}
\end{subequations}

We illustrate this result for an intermittent trap, $\mathcal{K}=\{0,\kappa\}$, switching between ``off'' and ``on'' states with rates $r_{\rm off}$ and $r_{\rm on}$. For symmetric switching, $r_{\rm on}=r_{\rm off}=r$, Eqs.~\eqref{eq:mom1}--\eqref{eq:mom2} yield
\begin{equation}
    \mathbb{E}_{\eta,k}[\dot{W}_{\rm steady}]
    =\frac{D\kappa}{2}+\frac{(\kappa+4r)v^2}{2r}.
    \label{eq:work_rate_constantv}
\end{equation}
Hence the dissipated work per unit distance, ${\mathbb{E}_{\eta,k}[\dot{W}_{\rm steady}]/}{v}$,
is non-monotonic in $v$, diverging as $v^{-1}$ for $v\to 0$ and as $v$ for $v\to \infty$. The minimum occurs at
\begin{equation}
    v^*=\argmin_v\left\{\frac{\mathbb{E}_{\eta,k}[\dot{W}_{\rm steady}]}{v}\right\}
    =\sqrt{\frac{D\kappa r}{\kappa+4r}}.
\end{equation}
Thus the dissipative nature of the controller generates a finite characteristic transport velocity already at the level of steady dragging. In the fast-switching limit $r\to\infty$, the positional dynamics approach those of Brownian motion in a static trap with stiffness $\bar{k}=\kappa/2$, while the thermodynamic cost remains anomalous \cite{bo2014entropy,celani2012anomalous}, i.e., positive, due to the persistent probe-controller energy exchange.

\begin{figure}
    \onefigure[width=\linewidth]{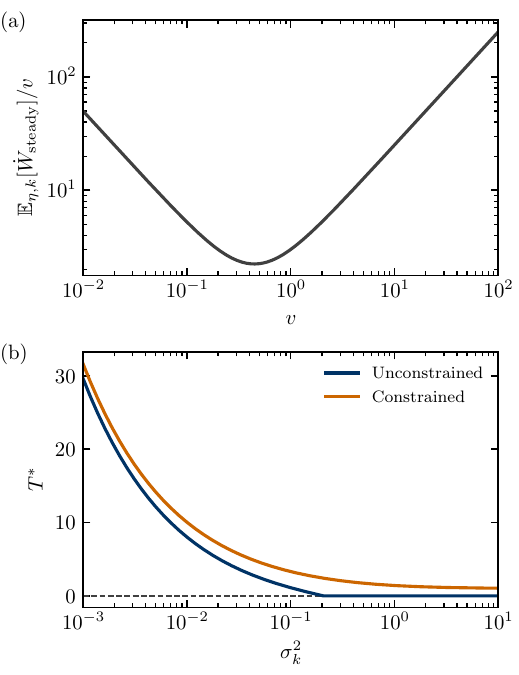}
    \caption{Optimal transport with a noisy trap under steady dragging. (a) Work per unit distance in an intermittent trap obtained from Eq.\,\eqref{eq:work_rate_constantv}. Here, $D=1$, $\kappa=1$ and $r=1$. (b) Optimal protocol duration for unconstrained [Eq.\,\eqref{eq:unconstrained_opT_transport}] and constrained [Eq.\,\eqref{eq:constrained_opT_transport}] optimal transport, showing the transition to zero-time optimality. Here, $D=1$, $\bar{k}=1$ and $\lambda_T=1$.}
    \label{fig:transport}
\end{figure}

\subsection{Optimal control}

We now return to the finite-time transport problem and seek the protocol $\lambda(t)$ connecting $\lambda(0)=0$ to $\lambda(T)=\lambda_T$ that minimises $\overline{W}[\lambda]$. To obtain analytical expressions, we consider the fast-switching regime by rescaling $\M\to \varepsilon^{-1}\M$ and sending $\varepsilon\to 0$ \cite{pavliotis2008multiscale}. In this limit, the stiffness dynamics are fast compared with the diffusive timescale of the probe. Expanding the joint distribution as $P_i=P_i^{(0)}+\varepsilon P_i^{(1)}+\mathcal{O}(\varepsilon^2)$ with $P_i^{(0)}=\pi_iP(u,w,t;k=\bar{k})$, and using Eq.~\eqref{eq:exp_invol_k} together with the Fokker-Planck equation \eqref{eq:FP_ot}, one finds $\langle ku\rangle=\bar{k}u+\mathcal{O}(\varepsilon)$ and
\begin{equation}
    \langle \dot{k}(\lambda^2+w-2\lambda u)\rangle
    =2\sigma_k^2(\lambda^2+w-2\lambda u)+\mathcal{O}(\varepsilon),
\end{equation}
so that the mean work reduces to
\begin{equation}\label{eq:w_OT_fast}
    \overline{W}[\lambda]
    =\int_0^T \dt\:
    \left[\bar{k}\dot{\lambda}(\lambda-u)+\sigma_k^2(\lambda^2+w-2\lambda u)\right].
\end{equation}
Note that this depends on $k(t)$ only through its mean and variance.
At the same order, the moment equations \eqref{eq:langevin_u_w} close as
\begin{subequations}\label{eq:fastslow_transport}
\begin{align}
    \dot{u} &= -\bar{k}(u-\lambda), \label{eq:fastslow_u} \\
    \dot{w} &= -2\bar{k}(w-\lambda u)+2D. \label{eq:fastslow_w}
\end{align}
\end{subequations}
Eliminating $\lambda$ through Eq.~\eqref{eq:fastslow_u} and enforcing Eq.~\eqref{eq:fastslow_w} with a Lagrange multiplier $\Lambda_w(t)$ gives
\begin{align}
    \overline{W}
    &=\left[\frac{\dot{u}^2}{2\bar{k}}\right]_{0}^{T}
    +\int_0^T \dt\:
    \Bigg[
        \left(1+\frac{\sigma_k^2}{\bar{k}^2}\right)\dot{u}^2
        +\sigma_k^2(w-u^2)\nonumber\\
        &\hspace{4.7em}
        +\Lambda_w\left(\dot{w}+2\bar{k}w-2u\dot{u}-2\bar{k}u^2-2D\right)
    \Bigg].
    \label{eq:work_avg}
\end{align}
Following the standard variational approach, the associated Euler-Lagrange equations are derived as
\begin{subequations}
\begin{align}
    0 &= -\left(1+\frac{\sigma_k^2}{\bar{k}^2}\right)\ddot{u}
    +u\left(\dot{\Lambda}_w-2\bar{k}\Lambda_w-\sigma_k^2\right),
    \label{eq:EL1_OT}\\
    0 &= \dot{\Lambda}_w-2\bar{k}\Lambda_w-\sigma_k^2.
    \label{eq:EL2_OT}
\end{align}
\end{subequations}
They are solved by the constant-speed trajectory
\begin{equation}
    u(t)=ct,\qquad w(t)=\frac{D}{\bar{k}}+c^2t^2,
\end{equation}
with $\Lambda_w(t)=\Lambda_0e^{2\bar{k}t}-\sigma_k^2/(2\bar{k})$, where $\Lambda_0$ is an integration constant that does not affect the physical trajectories. Substituting back into Eq.~\eqref{eq:work_avg} yields
\begin{equation}\label{eq:OT_work_c}
    \overline{W}(T;c)=\left(1+\frac{\sigma_k^2}{\bar{k}^2}\right)c^2T+\frac{\sigma_k^2D}{\bar{k}}T+\frac{\bar{k}}{2}(\lambda_T-cT)^2.
\end{equation}

For the \emph{unconstrained} problem, $c$ is free and minimisation of Eq.~\eqref{eq:OT_work_c} at fixed $T$ gives
\begin{equation}
    c^*(T)=\frac{\bar{k}\lambda_T}{2\left(1+\sigma_k^2/\bar{k}^2\right)+\bar{k}T}.
\end{equation}
Minimising the resulting $\overline{W}(T;c^*)$ over $T\geq 0$ gives
\begin{equation}
    T^*=
    \begin{cases}
        \lambda_T\sqrt{\dfrac{\sigma_k^2+\bar{k}^2}{D\sigma_k^2\bar{k}}}
        -\dfrac{2(\sigma_k^2+\bar{k}^2)}{\bar{k}^3},
        & \sigma_k^2<\sigma_{k,c}^2,\\[1.2ex]
        0, & \sigma_k^2\geq \sigma_{k,c}^2,
    \end{cases}
    \label{eq:unconstrained_opT_transport}
\end{equation}
where a critical variance emerges, which satisfies
\begin{equation}
    \sigma_{k,c}^2
    = \frac{\bar k^2}{2}\left( \sqrt{1+\frac{\bar{k}\lambda_T^2}{D}} -1 \right).
\end{equation}
Hence the optimal duration collapses to zero above a threshold in controller fluctuations; see Fig.~\ref{fig:transport}(b). This zero time optimal protocol signifies that the energetic cost of maintaining the dissipative controller for any length of time $T$ is not compensated for by the resulting reduction in dissipation in transporting the particle over time $T$ compared to the instantaneous case $T=0$.

This transition is captured cleanly by the small-$T$ expansion of the optimised work,
\begin{equation}
    \overline{W}(T;c^*)=\overline{W}(0;c^*)+\overline{W}'(0;c^*)\,T+\mathcal{O}(T^2),
\end{equation}
for which
\begin{equation}
    \overline{W}'(0;c^*)
    =\sigma_k^2\left[
    \frac{D}{\bar{k}}-\frac{\bar{k}^4\lambda_T^2}{4\sigma_k^2(\bar{k}^2+\sigma_k^2)}
    \right].
    \label{eq:smallT_transport}
\end{equation}
The coefficient changes sign precisely at $\sigma_k^2=\sigma_{k,c}^2$, marking the onset of local optimality of the zero-time protocol.

For the \emph{end-point-constrained} problem, we additionally require $u(T)=\lambda_T$, so that the mean probe position coincides with the final trap centre \cite{goerlich2025time,aurell2011optimal}. This fixes $c=\lambda_T/T$ and leads to
\begin{equation}
    \overline{W}_{\rm C}(T)=
    \left(1+\frac{\sigma_k^2}{\bar{k}^2}\right)\frac{\lambda_T^2}{T}
    +\frac{\sigma_k^2D}{\bar{k}}T,
\end{equation}
with optimum
\begin{equation}
    T^*_{\rm C}
    =\sqrt{\left(1+\frac{\sigma_k^2}{\bar{k}^2}\right)\frac{\lambda_T^2\bar{k}}{\sigma_k^2D}}.
    \label{eq:constrained_opT_transport}
\end{equation}
The constrained optimum therefore remains finite for all $\sigma_k^2\geq 0$ as shown in Fig.~\ref{fig:transport}(b). Similarly to the unconstrained case, $T^*$ tends to infinity as $\sigma_k^2\to 0$, as expected in the equilibrium limit.

\subsection{Extension to correlated probe noise}

The transport analysis can be extended to exponentially correlated probe noise with
\begin{equation}
    \mathbb{E}_\eta[\eta(t)\eta(t')]
    =\frac{e^{-|t-t'|/\tau}}{2\tau},
\end{equation}
where $\tau>0$ is the correlation time. This may be interpreted as replacing the equilibrium thermal bath by an active bath, e.g. a suspension of
motile particles \cite{wu2000particle,solon2015active}. The equation for $u$ is unchanged and given by Eq.\,\eqref{eq:fastslow_u}, while the variance evolves as
\begin{equation}
    \dot{w}=-2k(w-\lambda u)+\sqrt{8D}\,\mathbb{E}_\eta[x\eta],
\end{equation}
with
\begin{equation}\label{eq:w_noise_exp}
    \mathbb{E}_\eta[x\eta]
    =\int_0^t \dt'\:
    \sqrt{\frac{D}{2\tau^2}}
    \exp\!\left[-\frac{|t-t'|}{\tau}-\int_{t'}^{t}\dt''\,k(t'')\right].
\end{equation}
Assuming again fast stiffness fluctuations and finite $\sigma_k^2$, the averaged equation becomes
\begin{equation}
    \dot{w}
    =-2\bar{k}(w-\lambda u)
    +\frac{2D}{1+\bar{k}\tau}\left[1-e^{-(\bar{k}+\tau^{-1})t}\right].
\end{equation}
The source term now vanishes at short times $t$. As a result, while the Euler-Lagrange extremals remain of the form
\begin{subequations}
\begin{align}
    u(t)&=ct,\\
    w(t)&=c^2t^2+\int_0^t \dt'\:
    \frac{2D}{1+\bar{k}\tau}
    \left[1-e^{-(\bar{k}+\tau^{-1})t'}\right]e^{-2\bar{k}(t-t')},
\end{align}
\end{subequations}
the linear term in the small-$T$ expansion responsible for Eq.~\eqref{eq:smallT_transport} is no longer generated by the noisy-controller contribution. Consequently, $\overline{W}'(0;c^*)$ remains negative and the transition to zero-time optimal transport is suppressed. This confirms that the zero-time transition depends on the short-time structure of the stochastic forcing, with noncommuting fast-stiffess and fast-noise limits.

\section{Stiffening/softening trap}\label{sec:stiffening}

We now consider the second protocol class of Ref.~\cite{schmiedl2007optimal}, in which the trap centre is fixed and the mean stiffness is driven from $\lambda(0)=\lambda_0$ to $\lambda(T)=\lambda_T$. The probe obeys the Langevin dynamics
\begin{equation}
    \dot{x}(t)=-[\lambda(t)+k(t)]x(t)+\sqrt{2D}\,\eta(t),
\end{equation}
where the stochastic stiffness fluctuations have now zero mean, $\bar{k}=0$, and variance $\langle k^2\rangle=\sigma_k^2$. In the fast-switching regime, the variance of the probe position obeys
\begin{equation}
    \dot{w}=-2\lambda w+2D.
\end{equation}
Proceeding as in the transport problem, one finds from the first-order correction in the multiscale expansion that
\begin{equation}
    \langle \dot{k}\,w\rangle = 2\sigma_k^2w +\mathcal{O}(\varepsilon) ,
\end{equation}
and therefore the mean work reduces to
\begin{align}
    \overline{W}[\lambda]
    &=\int_0^T \dt\:
    \left(\frac{\dot{\lambda}w}{2}+\sigma_k^2 w\right)\nonumber\\
    &=\frac{1}{2}[\lambda w-D\ln w]_0^T
    +\int_0^T \dt\:
    \left(\frac{\dot{w}^2}{4w}+\sigma_k^2 w\right)~,
    \label{eq:w_stiff_fast}
\end{align}
which again depends on $k(t)$ only through its mean and variance.
The associated Euler-Lagrange equation,
\begin{equation}
    \dot{w}^2-2\ddot{w}\,w+4\sigma_k^2 w^2=0,
\end{equation}
has general solution
\begin{equation}
    w_c(t)=\frac{D\cosh^2[\sigma_k(t-2c)]}{\lambda_0\cosh^2(2c\sigma_k)},
    \label{eq:sol_EL_stiff}
\end{equation}
with $c$ an integration constant, after imposing the equilibrium initial condition $w(0)=D/\lambda_0$. Substituting Eq.~\eqref{eq:sol_EL_stiff} into Eq.~\eqref{eq:w_stiff_fast} gives
\begin{equation}
    \overline{W}(T;c)=\frac{1}{2}\left[f(T;c)-D+D\ln\!\left(\frac{D}{\lambda_0}\right)\right]+s(T;c),
\end{equation}
with
\begin{subequations}
\begin{align}
    f(T;c) &= \lambda_T w_c(T)-D\ln w_c(T),\\
    s(T;c) &= \frac{D\sigma_k}{\lambda_0}
    \frac{\sinh(\sigma_k T)\cosh[\sigma_k(T-4c)]}{\cosh^2(2c\sigma_k)}.
\end{align}
\end{subequations}

\begin{figure}
    \onefigure[width=\linewidth]{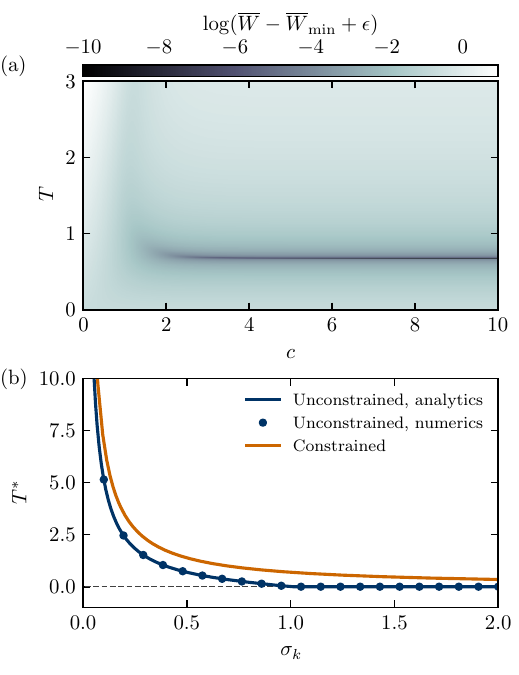}
    \caption{Optimal stiffening with a noisy trap. (a) Mean work as a function of the optimisation parameters $(c,T)$, showing convergence in the large-$c$ regime. (b) Optimal protocol duration for unconstrained [Eq.\,\eqref{eq:unconstrained_opT_stiff}] and constrained [Eq.\,\eqref{eq:constrained_opT_stiff}] optimal stiffening, showing the transition to zero-time optimality at $\sigma_{k,c} = (\lambda_T-\lambda_0)/2$. Here, $D=1$, $\lambda_0=1$, $\lambda_T=3$ and $\epsilon=10^{-10}$. 
    }
    \label{fig:stiffening}
\end{figure}

For the unconstrained problem, we minimise $\overline{W}(T;c)$ numerically over $(T,c)$; the mean work functional is shown in Fig.~\ref{fig:stiffening}(a). These numerics provide useful analytic guidance: we observe that the mean work functional becomes asymptotically independent of $c$, which means that the optimal protocol duration can be obtained by a simple one-dimensional minimisation with respect to $T$ in the large-$c$ limit. Using $\cosh x\sim e^x/2$ for $x\to\infty$, Eq.~\eqref{eq:sol_EL_stiff} becomes
\begin{equation}
    w_\infty(t)\sim \frac{D}{\lambda_0}e^{-2\sigma_k t},
\end{equation}
and the work functional approaches
\begin{equation}
    \overline{W}_\infty(T)
    =\frac{D}{2}\left(\frac{\lambda_T-2\sigma_k}{\lambda_0}\right)e^{-2\sigma_k T}
    +D\sigma_k T+\xi,
    \label{eq:mwf_final_stiff}
\end{equation}
with $\xi=D(\sigma_k/\lambda_0-1/2)$ independent of $T$. Minimising Eq.~\eqref{eq:mwf_final_stiff} with respect to $T$ yields the optimal protocol duration
\begin{equation}
    T^*=\frac{1}{2\sigma_k}\ln\!\left(\frac{\lambda_T-2\sigma_k}{\lambda_0}\right).
    \label{eq:unconstrained_opT_stiff}
\end{equation}
when $\sigma_k<\sigma_{k,c}= (\lambda_T-\lambda_0)/2$, and $T^*=0$ otherwise. This analytic result matches the $T^*$ found from the numerical minimisation, as seen in Fig.~\ref{fig:stiffening}(b). 

A small-$T$ expansion leads to writing 
\begin{equation}
\overline{W}_\infty(T) = \overline{W}_{\infty}(0) + \overline{W}'_{\infty}(0) T + \mathcal{O}(T^2)
\end{equation}
where
\begin{subequations}
\begin{align}
\overline{W}_{\infty}(0) &= \frac{D}{2}\left(\frac{\lambda_T}{\lambda_0}  - 1\right), \\
\overline{W}'_{\infty}(0) &= \frac{2 D\sigma_k}{\lambda_0}\left(\sigma_k-\frac{\lambda_T-\lambda_0}{2}\right).
\end{align}
\end{subequations}
Here again, we observe that the derivative of the optimised work at $T=0$ changes sign at the critical noise level $\sigma_{k,c}$ leading the unconstrained optimum collapsing to zero duration above this threshold as shown in Fig.~\ref{fig:stiffening}(b). 

For the end-point-constrained problem, we fix the final variance to its equilibrium value, $w(T)=D/\lambda_T$. Solving Eq.~\eqref{eq:sol_EL_stiff} for $c$ gives
\begin{equation}
    c(T)=\frac{1}{2\sigma_k}\operatorname{arctanh}\left[
    \frac{\cosh(\sigma_k T)-\sqrt{\lambda_0/\lambda_T}}{\sinh(\sigma_k T)}
    \right].
\end{equation}
A solution exists only for
\begin{equation}
    T>T_{\rm min}=\frac{\ln(\lambda_T/\lambda_0)}{2\sigma_k},
    \label{eq:constrained_opT_stiff}
\end{equation}
which is strictly positive in the stiffening case. Optimising numerically over the admissible range then yields $T^*=T_{\rm min}$ for all $\sigma_k^2$ (Fig.~\ref{fig:stiffening}b). As in the transport scenario, the additional end-point constraint removes the transition to zero-time optimality.

\section{Conclusion}

We have studied finite-time control of a passive Brownian probe by a noisy harmonic trap whose stiffness dynamics are dissipative as a consequence of violating local detailed balance. In both transport and stiffening protocols, the nonequilibrium probe-controller coupling qualitatively alters the control landscape: the optimal duration is finite even though the controlled particle remains passive, and the unconstrained optimum can collapse to zero duration above a critical level of trap fluctuations. This transition follows directly from the short-time expansion of the optimised work: the noisy controller generates a positive contribution linear in $T$, which competes with the negative short-time contribution associated with deterministic control, for which slower protocols are generically more efficient; the cancellation of the corresponding prefactors marks the onset of local optimality of the zero-time protocol. This is the appropriate small-$T$ counterpart, in the present setting, of the broader scaling arguments used elsewhere to rationalise finite-time optima in active systems \cite{davis2024active,soriani2025control} and passive systems subject to slow noisy control \cite{large2012stochastic}.
Although we formulate the problem here for $k(t)$ being a discrete-state process, we expect the same results to hold for continuous stochastic fluctuations of the trap stiffness, and have verified this explicitly for the transport problem with Ornstein–Uhlenbeck stiffness fluctuations (the so-called OU$^2$ process \cite{cocconi2024ou2}).

Note that, strictly speaking, the fast-slow limit employed in the derivation of these results only holds for finite $T$, i.e.\ for $\sigma_{k}^2$ arbitrarily close (but not equal to) its critical value. Going beyond leading order in the multiscale expansion would allow for a more refined description of the scaling behaviour of $\overline{W}$ as $T$ approaches zero. We leave this technical challenge for future work.

The constrained problems behave differently. Imposing a final-state condition removes the zero-time transition both for transport and stiffening, leaving a strictly positive optimal duration. This makes clear that the transition is not simply a consequence of controller dissipation, but of how that dissipation competes with the admissible control manifold at short times. The extension to correlated probe noise sharpens this point further: finite-time optimality persists, while the zero-time transition is suppressed when the nonequilibrium forcing no longer contributes at order $T$, highlighting the noncommutativity of the fast-stiffness and fast-noise limits.

More broadly, our results identify a generic route to finite-time optimality in passive systems controlled by nonequilibrium devices. This mechanism should be relevant whenever the controller continuously exchanges work and heat with the system during the protocol, for instance in fluctuating optical traps or in complex active environments coupled to passive degrees of freedom. \\

\section{Acknowledgements} We thank Sarah Loos and Rémi Goerlich for interesting discussions. LC and HA gratefully acknowledge support from the International Center for Mechanical Sciences (CISM) where part of the work presented here was conducted during a short residence. LC acknowledges funding from the Ernest Oppenheimer Fund.

\bibliographystyle{eplbib}
\bibliography{references}

\end{document}